\documentclass[twocolumn,showpacs,preprintnumbers,amsmath,amssymb]{revtex4}

\usepackage{graphicx}
\usepackage{dcolumn}
\usepackage{bm}
\usepackage{natbib}

\begin{document}


\title{Long-term stabilization of the length of an optical
reference cavity}

\author{Ga\"etan Hagel}
\author{Marie Houssin}
\author{Martina Knoop}  \email{Martina.Knoop@up.univ-mrs.fr}
\author{Caroline Champenois}
\author{Michel Vedel}%
\author{Fernande Vedel}

\affiliation{Physique des Interactions Ioniques et Mol\'eculaires
(CNRS UMR 6633), Universit\'e de Provence, Centre de Saint
J\'er\^ome, Case C21, 13397 Marseille Cedex 20, France}%
 \homepage{http://www.up.univ-mrs.fr/ciml/}

\date{\today}

\begin{abstract}
To obtain a high degree of long-term length stabilisation of an
optical reference cavity, its free-spectral range is locked by
means of an accurate and stable frequency synthesizer. The locking
scheme is twofold: a laser is locked on the N$^{th}$ mode of a
reference Fabry-Perot cavity and part of the laser light is
shifted in frequency  to be in resonance with the (N+1)$^{th}$
mode of the cavity. This shift   is generated by an
acousto-optical modulator (AOM) mounted in a double-pass scheme,
matching half of the free spectral range of the reference cavity.
The resulting absolute stabilization of the length of the cavity
reaches the 10$^{-11}$ level per second, limited by the lock
transfer properties and the frequency stability of the AOM control
synthesizer.
\end{abstract}

\maketitle

\section{Introduction}
In recent years, optical frequency metrology has made spectacular
progress with the implementation of femtosecond frequency combs
\cite{udem02}.
 Today, the precision and stability
of laser frequencies reaches beyond the 10$^{-14}$-level, removing
the last obstacles for an ultimate investigation of possible new
frequency standards outpassing the performances of the actual
cesium atomic clock by orders of magnitude \cite{diddams01}.
 Nevertheless, many experimental applications
require lasers with less stringent performances, like  a sub-MHz
linewidth and an absolute frequency stability better than this
spectral width for time scales of a minute. In our experiment
\cite{lisowski05b}, probing of an atomic transition requires a
narrow interrogation laser linewidth, which can be obtained in a
straightforward way by stabilizing a laser on a Fabry-Perot cavity
of high finesse. The use of the Pound-Drever-Hall (PDH) locking
technique \cite{drever83}  permits to counteract rapid frequency
fluctuations and allows to obtain laser linewidths largely
inferior to 100~kHz.

To reach integration times of a couple of minutes, frequency
drifts have to be suppressed, which requires  stabilisation of the
length of the reference cavity  of the interrogation laser. This
can be realized by a maximum isolation from mechanical, acoustic
and thermal perturbations, or by an absolute stabilisation  on an
atomic transition, generally by making use of an additional
(diode) laser set-up. Stabilisation on an atomic transition is
often made by the saturated absorption technique to avoid Doppler
broadening effects in a gas at room temperature. The employed
crossover  transitions have spectral widths of about 10~MHz (for
example Rb: 5.9~MHz, Cs: 14~MHz), limiting the attainable
frequency stabilities to almost three orders of magnitude below
these values.

Very few different techniques have been used to assure an
absolute-length stabilisation of a reference cavity. They all
compare the frequency difference between two eigenmodes of a
reference cavity with an external rf frequency applied to a phase
or frequency modulation device. The Dual-Frequency Modulation
(DFM) technique with two phase modulators (EOM) in a row has  been
used in \cite{deVoe84}. It has been applied in a slightly modified
set-up to measure the frequency of molecular transitions with a
 uncertainty of 2 $\times 10^{-8}$ \cite{ma99}, where the
change in free-spectral range (FSR) of an optical cavity is
tracked by a servo on an EOM. Another implementation is reduced to
the use of a single EOM \cite{courteille94}. Long-baseline
interferometry uses double modulation techniques to assure length
determination of the interferometer arms which can reach a
relative uncertainty of 10$^{-11}$ \cite{araya99}.

In this manuscript we describe a new scheme for the use of a DFM
method for the absolute-length stabilisation of an optical cavity
for measurements on timescales reaching from several seconds up to
hours. In order to avoid a long-term drift of the laser frequency
we have chosen to stabilize the length of the reference cavity by
locking the frequency difference between its N$^{th}$ and its
(N+1)$^{th}$ longitudinal mode. The frequency difference is
generated by an acousto-optical modulator at a fixed value
matching  half of the FSR of the reference cavity. The high
accuracy and stability of a frequency synthesizer are thus
transferred  to the FSR. The use of acousto-optic modulators at
frequencies largely beyond 100~MHz allows to stabilize short
optical cavities. Furthermore, the described method is independent
from the existence of nearby atomic transitions. We have
demonstrated excellent frequency stability for periods up to an
hour.

\section{Experimental realization}

\begin{figure}
\includegraphics[width=90mm]{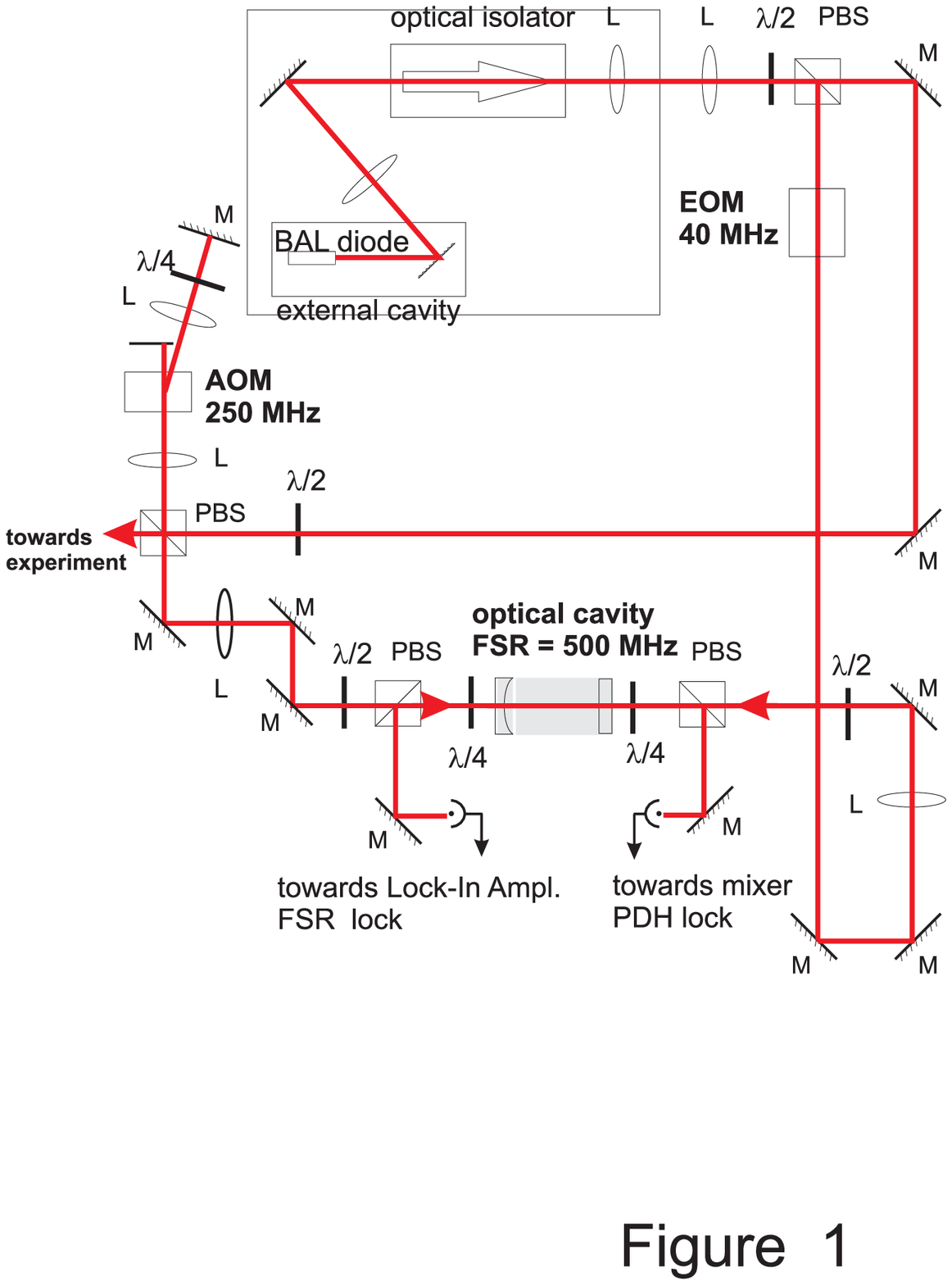}
\caption{Experimental setup of the described locking scheme. The
 signal for the PDH lock of the diode laser enters the
cavity from the right, while the signal for the length
stabilisation of the cavity is injected into the reference cavity
from the left side, after a double-pass through the 250-MHz AOM. L
designs the lenses used for efficient mode-matching, while mirrors
are noted M, and polarization beam-splitters PBS. See text for
detailed description.} \label{fig:setup}\end{figure}

The complete experimental set-up is shown in figure
\ref{fig:setup}. The 729-nm-laser  is a broad-area diode laser
(BAL) mounted in an external cavity to assure single-mode lasing
and to reduce its linewidth \cite{houssin03}. The diode is then
stabilized onto a reference cavity by using the Pound-Drever-Hall
lock \cite{drever83}. The laser light is phase-modulated at 40 MHz
by an EOM. The signal reflected by the cavity is composed of two
totally reflected sidebands and a central carrier whose phase
depends on the frequency difference with a cavity resonance.
Detected by a rapid photodiode, the beat signals between the
carrier and the sidebands allow to generate an error signal that
permits frequency corrections up to the cut-off frequency of the
cavity and phase corrections beyond.


Our present goal is to enhance the long-term stability of the
optical reference cavity. The cavity has a finesse of about 1000
and the spacer between the two mirrors is made out of Invar. This
home-made cavity spacer has an outer diameter of 50~mm, an inner
diameter of 10~mm, and a length of about 300~mm. The cavity is
formed by commercial broad-band mirrors (700-900~nm), a plane and
a curved one (radius of curvature 2m). The plane mirror is mounted
on a piezo-electric transducer (PZT) allowing to adjust the length
of the cavity by application of a DC-voltage from 0~V up to 150~V
corresponding to a maximum length variation of about 2~$\mu$m. In
the course of a day, the reference cavity undergoes a slow
frequency drift mainly due to temperature variations, in a lab
where the ambient temperature is stabilized to approximately one
degree.

\begin{figure}
\includegraphics[width=90mm]{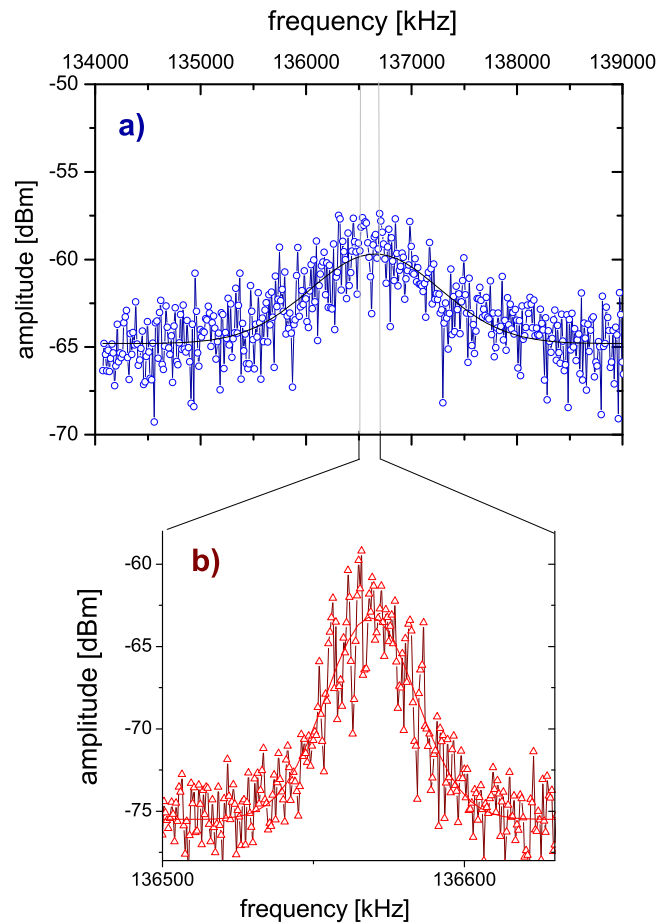}
\caption{Beat signal of one part of the 729~nm diode laser with
another part of the same laser having passed through a 10~km
optical fiber. Part a) shows a recording  for the free-running
diode resulting in a spectral linewidth of several hundred kHz,
figure b) has been recorded with the diode laser being locked to
the reference cavity, and results in a spectral linewidth lower
than 25 kHz. }\label{fig:larg729}
\end{figure}

The instantaneous linewidth of the diode laser locked onto the
reference cavity has been measured by an autocorrelation technique
\cite{okoschi80}. The beat signal of a part of the laser beam with
a second part of the same beam which has passed through 10~km of
optical fiber is shown in figure \ref{fig:larg729}. This beat
signal has been acquired with a HP ESA 1500A spectrum analyzer at
a 3~kHz bandwidth. While the free-running laser (figure
\ref{fig:larg729} a) ) has a linewidth of the order of several
hundred kHz, which is a typical result for a broad-area diode
laser in an external cavity, the linewidth of the locked laser is
lower than 25 kHz (figure \ref{fig:larg729} b) ). This is achieved
by feeding the frequency corrections of the PDH-signal to
different elements in the diode laser cavity setup. Slow
fluctuations (up to 200 Hz) are corrected by the piezo-electric
element supporting the cavity grating with a resistor-bias
stabilization integrator (pseudo-integrator) \cite{Horowitz89}.
Intermediate corrections (up to 20 kHz) are applied via the diode
laser current source, and rapid frequency corrections are made
directly on the anode of the diode laser making use of a FET
mounted as a passive voltage-current convertor
\cite{schmidtkaler03pr}. The difference in gain between slow and
rapid corrections is approximately 20~dB. The total bandwidth of
the servo loop is about 1~MHz.

To increase the stability of the reference cavity in order to
avoid a long-term drift of the laser we have chosen to stabilize
the length of the cavity by locking the frequency difference
between its N$^{th}$ and its (N+1)$^{th}$ longitudinal mode. The
frequency difference is fixed by a highly stable frequency
synthesizer driving the acousto-optical modulator at a frequency
corresponding to half of the cavity FSR. We use an Aeroflex
2030-series frequency synthesizer  with an internal frequency
standard at 10 MHz (OCXO) having a 0.1~Hz accuracy  and a
fractional frequency stability at 1 minute of about 5 $\times
10^{-10}$ \cite{marconi}.

In the described locking scheme, the frequency of the diode laser
corresponds to the N$^{th}$ multiple of the cavity's free spectral
range \cite{layer76}
\begin{equation}
\nu_{DL} = ( N + \varphi + \Phi )\cdot \nu_{FSR}
\end{equation}
where $N$ is the mode number, $\varphi$  the Fresnel phase shift
due to the curvature of the cavity mirrors, and $\Phi$   the phase
shift which occurs upon reflection due to the finite conductivity
of the mirrors. In the following, we are only interested in the
frequency difference of two neighboring modes, we may thus assume
that these phase shifts are almost identical and that their
differences can therefore be neglected.

In addition to the PDH lock of the diode onto the reference
cavity, one part of the laser output which has double-passed an
acousto-optic modulator is injected into the cavity. Note that in
our set-up this second beam enters the cavity from the opposite
side of the PDH lock.

This beam has been offset in frequency by two times $\nu_{AOM}$ =
249.82~MHz fixing the FSR of the reference cavity.
\begin{equation}
\nu_{DL} + 2 \times \nu_{AOM} = ( N +1 ) \cdot \nu_{FSR}
\end{equation}

For the lock loop the driving frequency of the AOM is modulated at
$f_m$=22~kHz with a 100~kHz amplitude corresponding to 6~\% of the
width of the Airy peak of the cavity. The light reflected by the
cavity is then collected by a photodiode and demodulated at 22~kHz
by a lock-in amplifier. The output of this lock-in is integrated
before being sent to  the piezo-electric transducer controlling
the length of the reference cavity. For long-term corrections a
time constant of 1 second has been chosen.

In the present optical set-up  the two  photodiodes generating the
error signals are sensitive to both parts of the beam.   To avoid
crosstalk of the photodiodes in response of both beams, we have
separated the counter-reaction by choosing different bandwidths
for the two lock loops. In practice, our locking electronics has
been realized with a pure integrator in the length-stabilisation
of the reference cavity and a pseudo-integrator in the PDH lock
limiting the gain at zero frequency. As a consequence, at
frequencies below the hertz, the corrections applied to the diode
laser are negligible compared to those sent to the FSR lock.

The separation of the locks by their  bandwidths is the most
simple solution, as both loops work on different timescales. In a
more general scheme, the lock loops could be separated by making
use of orthogonal polarizations. However, the optical separation
should be made with Glan-Laser prisms as the polarization
beam-splitter cubes used in our set-up present an extinction ratio
of only 0.01.

A second essential point for a stable configuration of the locking
scheme is to choose a frequency modulation  $f_m$ of the FSR lock
larger than the cut-off of the current corrections in the
diode-lock ($\approx$ 2 kHz). At small frequency modulation values
(e.g. 1 kHz) the gain of the PDH servo is  high (55dB for the
electronic part) producing a strong reaction by correcting
modulation as noise. As a consequence the lock of the laser on the
reference cavity becomes unstable. For best results, we have
chosen to modulate the driving frequency of the AOM  at 22 kHz
with an amplitude of 100 kHz. At this frequency the electronic PDH
gain is about 35 dB, and the modulation does not perturb the PDH
lock.

The overall response of the lock is given by the resolution of the
employed frequency synthesizer. In fact,
\begin{equation}
\frac{\Delta\nu_{DL}}{\nu_{DL}} =
\frac{\Delta\nu_{FSR}}{\nu_{FSR}} =
\frac{\Delta\nu_{synth}}{\nu_{synth}}
\end{equation}
and the ratio of the FSR to diode frequency is, in our case,
\begin{equation}
\nu_{DL} \approx 0.8 \times 10^6 \nu_{FSR} = 1.6 \times 10^6
\nu_{synth}
\end{equation}

\begin{figure}
\includegraphics[width=90mm]{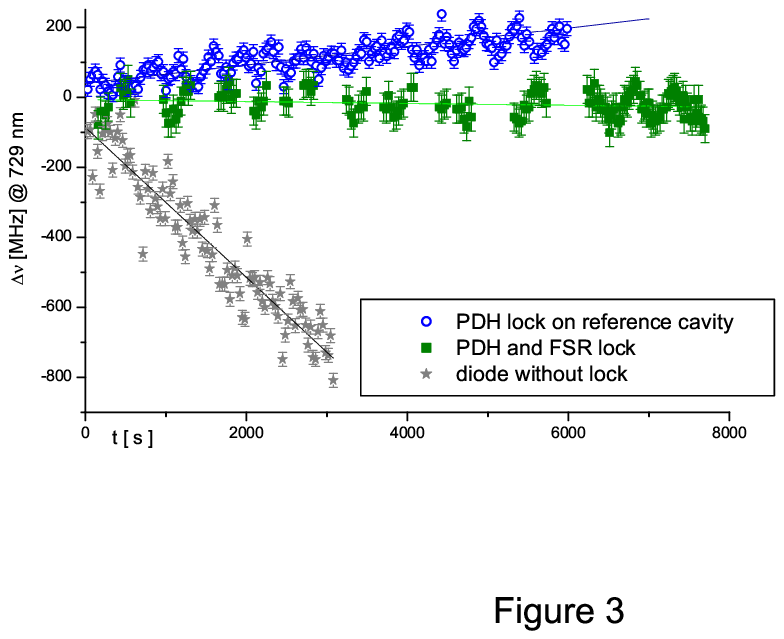}
\caption{Temporal evolution of the laser frequency for a
free-running laser ($\star$), the laser locked by PDH method on
the optical cavity ($\circ$), and with the FSR of the cavity
locked on the frequency synthesizer ($\bullet$). }
\label{fig:evolp}\end{figure}

The time constant of the locking electronics is about one second
to assure a high long-term stability of the system, the timescale
of interest for our experiment is of the order of a couple of
minutes. To test the achieved stability, we have therefore
measured the evolution of the diode laser frequency every 30
seconds making use of  a wavemeter, where the employed reference
wavelength is a temperature stabilized HeNe transition at
632.8~nm. The absolute precision of the wavemeter is better than
40 MHz, which has been verified by the frequency resolution
obtained on the atomic transition of a single calcium ion during
periods of a couple of hours \cite{knoop04}. The temporal
evolution of the diode laser frequency is reported in figure
\ref{fig:evolp}, the straight lines in this graph are linear fits
to the acquired data, reflecting the long-term frequency drifts.
On the considered time scales (1 minute to three hours) the
observed frequency variations have all presented a linear
evolution.

The lower curve in figure \ref{fig:evolp}  reflects the frequency
of the diode laser in external cavity without any additional
stabilisation. The frequency drift is almost 800 MHz per hour,
mainly due to thermal drifts of the high-power diode-laser
component, that is a very typical value for a non-stabilized
laser. The upper curve shows the fluctuations of the diode
frequency as it is stabilized onto the described Invar cavity. A
reduction of the frequency drift to values below 100 MHz per hour
can be observed. The center graph monitors the frequency of the
diode locked to the reference cavity with  the cavity's FSR
stabilized by the synthesizer frequency. The apparent drift of the
frequency has been reduced to less than 3~kHz per second,
corresponding to a frequency variation per hour of less than 9 MHz
limited by the frequency stability of the employed frequency
synthesizer.

\section{Conclusion}
We have presented a method to stabilize the length of an optical
reference cavity on a long timescale implementing the lock of its
FSR by an AOM. The corner stone of this method is a thorough
frequency separation  of the two lock loops, the result is
dependant of the quality of the locking schemes adopted. The
stability reached by this method is only limited by the technical
performance of the employed frequency synthesizer. The implemented
technique is a pure frequency lock, no phase condition has to be
fulfilled. Compared to existing DFM methods which have been
designed using phase modulators, the use of an acousto-optical
modulator in our scheme allows application of the technique to
cavities which are shorter than some tens of centimeters.
Furthermore, the presented method can be used in a wavelength
regime where no atomic transitions are accessible. The stability
performances described could be easily improved by using an
external frequency standard with higher precision and stability
for the frequency synthesizer fixing the length of the optical
cavity.

\newpage

\end{document}